\documentclass[12pt]{article}

\usepackage{scicite}

\usepackage{times}

\usepackage{amsmath, amssymb, physics, bm}
\usepackage[a4paper,margin=1in]{geometry}
\usepackage{graphicx} 
\usepackage{xcolor}

\usepackage[a4paper,margin=1in]{geometry}

\newenvironment{sciabstract}{%
\begin{quote} \bf}
{\end{quote}}

\newcounter{lastnote}

\title{An electromagnetic model of the electron}

\author
{Carlos A. M. dos Santos$^{1,a}$, Marc J. J. Fleury$^{2,b}$\\
\\
\normalsize{$^{1}$Universidade de S{\~a}o Paulo, Escola de Engenharia de Lorena, Lorena-SP, 12.602-810, Brazil,}\\
\normalsize{$^{2}$Department of Physics and Astronomy, University of Nebraska-Lincoln, Lincoln, NE 68588, USA}\\
\\
\normalsize{$^{a,b}$To whom correspondence should be addressed; E-mail: cams-eel@usp.br, mfleury4@unl.edu}
}

\date{}

\begin{document} 


\baselineskip24pt


\maketitle


\begin{sciabstract}
We present a toroidal electromagnetic ansatz that provides a realistic microscopic model of the QED electron. The proposed toroidal electromagnetic wave satisfies Maxwell's equations and reproduces fundamental properties of the electron as described in quantum electrodynamics (QED). Within this framework, the electron is modeled as a rotating electromagnetic wave confined to a toroidal geometry. Parameter optimization yields quantitative agreement with the electron charge e, spin $\hbar/2$, and magnetic moment $\mu_B(1 + \alpha/2\pi)$, incorporating the Schwinger anomalous magnetic moment correction. The model yields an amplitude on the order of the Schwinger scale where electron-positron pair production occurs. The major radius corresponds to the Compton wavelength scale, while the monochromatic frequency is consistent with the de Broglie-Dirac frequency. The phase velocity is found to be $2c$, and the computed rest energy approximates $0.8 m_e c^2$. This representation provides a microscopic classical electromagnetic framework that encapsulates the properties of the QED electron.
\end{sciabstract}


\section*{Introduction}
The electron's nature remains fundamental to understanding physics, as both Wilczek and Einstein noted: "to understand the electron is to understand the world." While the electron's properties—charge, mass, spin, and magnetic moment—are precisely measured experimentally, the Standard Model describes it as a structureless elementary point particle. There is no realistic microscopic model of the electron. 

\section*{Historical Background}

The search for a physical microscopic model of the electron has its roots in the nineteenth-century hydrodynamic conception of matter and aether. In 1867, William Thomson (Lord Kelvin) proposed that atoms were vortex rings or knotted tubes of motion in an ideal, incompressible aether, whose stability and identity arose from their topological linkage~\cite{Kelvin1867}. This vortex-atom hypothesis inspired later efforts to describe charge and mass in terms of continuous fields rather than point particles. By the late nineteenth century, J.~J.~Thomson, H.~A.~Lorentz, and M.~Abraham had reformulated Kelvin’s idea in electromagnetic terms, treating the electron as a localized region of electromagnetic momentum or as a charged sphere whose inertia was of purely electromagnetic origin~\cite{Thomson1904,Lorentz1909}. With the advent of relativity and quantum mechanics, these classical models gave way to field-theoretic interpretations of the electron, yet the notion of topological structure in the field persisted. A century after Kelvin, Rañada showed that Maxwell’s equations admit exact, finite-energy knotted field solutions—Hopfions—in which electromagnetic helicity is a conserved topological charge \cite{Ranada1989,Ranada1990}. Building on this, Irvine and collaborators later produced knotted light fields experimentally from superposed Laguerre–Gaussian modes \cite{Irvine2008,Kedia2013}. All such configurations are, however, divergence-free and therefore carry no electric charge.

There is also a long history of attempts at providing a realist interpretation for the abstract objects encoded in the relativistic Dirac equation with point charges.  Such notable efforts can be traced back to Dirac himself, realizing his equations described what looked like an oscillation at the speed of light internal to the electron, to Schrodinger who then coined the term Zitterbewegung, literally jittery motion. This lineage of the electron considered as a charged field continued to the "classical model of the Dirac Electron" by Barut and Zanghi \cite{BarutZanghi1984}, to the vortex cores of Bohm, Vigier and Lochak \cite{BohmVigier} and the "Geometric Clifford Algebras" of Hestenes \cite{Hestenes1984}\cite{Hestenes2003}, but to single a few.  These frameworks usually remain rather abstract in their formalism but all roughly posit that the electron contains a charge and that far from being a abstract point has an internal 3D structure where this charge moves (usually at c) in a given geometry.

A critical review of recent zitterbewegung models by one author (MF) found existing approaches, particularly those invoking point-particle concepts, even within the extended geometries, to be inadequate \cite{FleuryRousselle2025}. That analysis identified three-dimensional electromagnetic wave-fields with zitterbewegung geometry as a promising research direction, combining Kelvin's approach with geometric insights. This framework, previously proposed by the first author (CdS) and reviewed in \cite{FleuryRousselle2025}, motivated our present deeper investigation of toroidal fields possessing divergence and thus intrinsic charge.
\section{Model}
\subsection{Properties of the electron}
Any model of the electron must account for the following observed properties:
\begin{itemize}
\item Compton scale dimensions ($\sim10^{-13}$ m) from experimental observations
\item Spin $S=\hbar/2$
\item magnetic moment $\mu=\mu_B=e\hbar/2m$, and $g$-factor $g=-2$ from Dirac theory
\item Anomalous $g$-factor: CODATA value $g=-2.00231930436256(35)$
\item Origin of electric charge
\item Mass and energy values
\end{itemize}
\subsection{Methodology}
We employ the following methodology: we postulate an electromagnetic ansatz specifying fields $\vec{E}$ and $\vec{B}$ using toroidal geometries consistent with the zitterbewegung hypothesis. We verify compliance with Maxwell's equations, deriving charge density as a geometrical property. Upon confirming the ansatz represents a valid electromagnetic wave, we derive QED electron phenomenology through parameter fitting. Failure to match the full phenomenology invalidates the ansatz. Our specific ansatz reproduces all target values with energy at $0.8m_ec^2$. While constituting a negative result, we believe this merits attention as it demonstrates that even a crude ansatz can account for most quantum properties. Detailed derivations appear in the appendix, providing systematic and transparent calculations.

\subsection{Electromagnetic Ansatz}
\subsubsection{Definitions}

\begin{figure}
\includegraphics[width=15 cm]{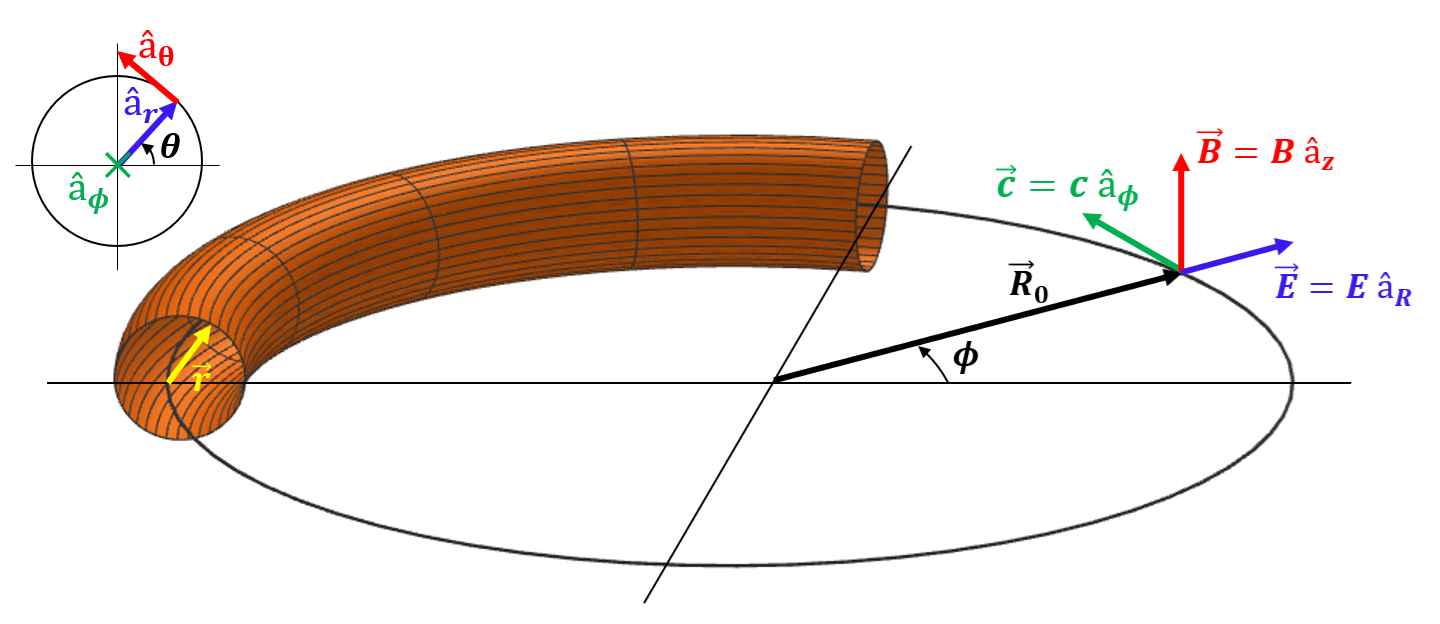}
\centering
\caption{The fields are defined inside the torus. The mask will be the step function effectively selecting the open inner volume as the domain where the EM field is defined. $R_0$ is the major radius, $r_0$ the minor radius. $\vec{R}$ is the position from origin to a given point inside the torus with $\vec{R} = \vec{R_0}+\vec{r}$}
\label{toursDef}
\end{figure}

Following the zitterbewegung intuition, we posit that the electron is an electromagnetic wave confined to a torus (see fig:\ref{toursDef}). We define the amplitudes as $E_0$ and $B_0=E_0/c$, but we do not specify their values. We call the major radius $R_{0}$ and the little radius $r_0$. We will define a vortex phase $\psi = (\phi - wt)$ but will not specify the frequency.  All four parameters, the amplitude, the major and minor radius of the torus and the frequency are free parameters of the model.  We will consider that the fields only exist within the volume of the torus and are null outside of it. We will make use of the Heaviside step function mask to quantify this.

Throughout the derivation, and for convenience, for vectors we will use the cylindrical coordinate basis $X=(R,\phi,z) $ and for integrations we will make use of the toroidal coordinate system to integrate over the volume delineated by the torus volume as shown in fig. 1.

\begin{figure}
\includegraphics[width=15 cm]{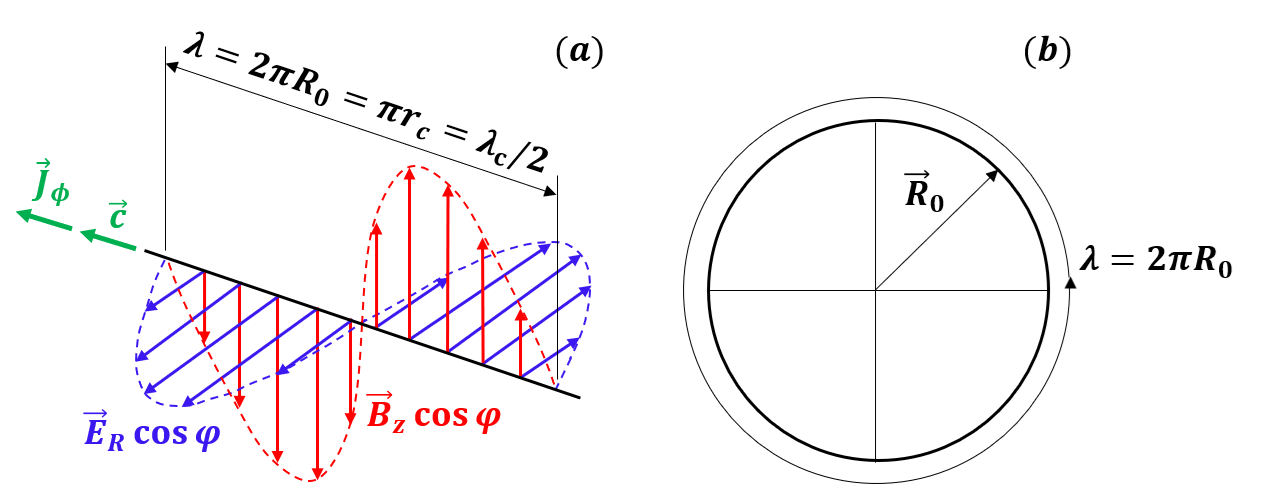}
\centering
\caption{E and B fields. Note the left handed nature. While the fields are represented flat in the (a) picture, they should be imagined as laid out along the circle on (b).  There is one wavelength fitting in the circle given the phase $\Psi= (\phi - \omega t)$.}
\label{EB}
\end{figure}   

\subsubsection{E and B}
We posit the following ansatz for the $\vec{E}$ and $\vec{B}$ electro-magnetic vectors

\begin{equation}
\vec{E} = iE_0 H({r}-{r}_0) e^{i(\phi-\omega t)}\!\left[
\hat{a}_R + i\!\left(1+\frac{R}{R_0}\right)\!\hat{a}_\phi
\right].
\end{equation}

\begin{equation}
\vec{B} = iB_0 H({r}-{r}_0) e^{i(\phi-\omega t)}\!
 \, \hat{a}_z.
\end{equation}

where
$E_0$ is the amplitude of the wave which we consider a free parameter. $B_{0} = {E_{0}}/{c}$, and $H({r}-{r}_0) = 1$ for $r < r_0$ and $0$ for $r > r_{0}$ is the Heaviside function.

$E_0, R_0, r_0, \omega$ are respectively amplitude, major and minor radius of torus and frequency. None of these values are specified for now and they treated as free parameters of the model.  Parameter fitting will set their values. 
See fig\ref{toursDef} and \ref{EB}. 


This ansatz exhibits an unconventional structure distinct from plane wave solutions to Maxwell's equations. The wave follows circular geometry with vectors aligned to cylindrical coordinates: $\vec{E}$ has components along both the radial direction $\hat{a}_R$ and the azimuthal direction $\hat{a}_\phi$, introducing a longitudinal component absent in purely transverse plane waves. The magnetic field $\vec{B}$ points along the $z$-direction. The phase $\theta = (\phi - \omega t)$ follows the classical phase vortex formalism \cite{Ceperley1992}. Notably, $\vec{E}$, $\vec{B}$, and the phase propagation direction form a left-handed triad, contrasting with the right-handed orientation of conventional electromagnetic plane waves.

We consider a non-planar monochromatic wave propagating circularly, confined to a torus, with angular frequency $\omega$. This represents a novel class of Maxwell solutions not previously documented. We now demonstrate that this ansatz satisfies Maxwell's equations.

\section{Maxwell's equations}

\subsection{Gauss' law for magnetic fields}

Divergence of $\vec{B}$ in cylindrical coordinates 

\begin{equation}
\vec{\nabla} \cdot \vec{B} =
\!\left\{\frac{1}{R}
\pdv{R}(R\, B_R) +
\frac{1}{R}\pdv{\phi}(B_\phi) +
\pdv{z}(B_z)
\right\}.
\end{equation}
Since $B_{R} = 0, B_{\phi} = 0$ their divergence is null. Since $B_{z} = i B_{0} e^{i(\phi-\omega t)}$, it obviously does not depend on $z$ and therefore it's derivative with respect to $z$ is also trivially null. Therefore, we obtain the trivial result that $\nabla \cdot \vec{B} = 0.$, e.g. no magnetic monopoles.

\subsection{Gauss' law for $\vec{E}$} 

In this section we will focus on Gauss' law for electric fields: 

\begin{equation}
\label{eq:gauss}
\vec{\nabla} \cdot \vec{E} = \frac{\rho}{\varepsilon_0},\qquad
\end{equation}

We employ the field-theoretic concept of geometrical charge. In conventional electrical engineering, the causal narrative proceeds from charges to fields: one introduces charge density $\rho$, induces motion, and generates radiating fields. This interpretation reads equation \ref{eq:gauss} from right to left—charges create fields. Here, charges are ontologically fundamental while fields emerge from their presence and dynamics.

We invite the reader to reverse the conventional reading of equation \ref{eq:gauss}, interpreting it from left to right: fields generate charge. This simple reversal emphasizes our inverted causality—we begin with a field ansatz and compute its divergence, identifying this as geometrical charge. Any non-zero divergence constitutes the charge density, a scalar quantity that merely quantifies the field's divergence without independent ontological status. In this framework, only fields possess objective reality; charge density emerges as an abstract, derived quantity.

Divergence of $\vec{E}$ in cylindrical toroidal coordinates
\begin{equation}
\vec{\nabla} \cdot \vec{E} =
\left\{\frac{1}{R}
\pdv{R}(R\, E_R) +
\frac{1}{R}\pdv{\phi}(E_\phi) +
\pdv{z}(E_z)
\right\}.
\end{equation}

The Electric field components are
\begin{equation}
E_R = iE_0 e^{i(\phi-\omega t)},\\
E_\phi = -\,E_0 e^{i(\phi-\omega t)}\!\left(1+\frac{R}{R_0}\right),\\
E_z = 0.
\end{equation}

which yields a divergence 
\begin{equation}
\vec{\nabla} \cdot \vec{E} = -\,i\frac{E_0}{R_0} e^{i(\phi-\omega t)},\\
\end{equation}

so we then define the charge density as
\begin{equation}
\rho(\phi,t) = -\,i\varepsilon_0 \frac{E_0}{R_0} e^{i(\phi-\omega t)}.
\end{equation}

Note that at this stage we have identified the charge with the divergence. We \textit{define} the charge density as the divergence of the $\vec{E}$ field. 

\subsection {Faraday's law}

We have specified the forms of the fields $\vec{E}$ and $\vec{B}$, and they must verify Faraday's law:
\begin{equation}
\vec{\nabla} \times \vec{E} = -\,\pdv{\vec{B}}{t}
\end{equation}

We will use this equation to set the frequency $\omega$, which is a free parameter. 

Resulting in
\begin{equation}
\vec{\nabla} \times \vec{E} = (\vec{\nabla} \times \vec{E})_z = - 2\,\frac{E_0}{R_0}\,e^{i(\phi-\omega t)}\,\hat{a}_z.
\end{equation}

Partial time derivative of $\vec{B}$ provides

\begin{equation}
\vec{B} = iB_0 e^{i(\phi-\omega t)}\,\hat{a}_z,\\
\pdv{\vec{B}}{t} = -\,i\omega \vec{B} = \omega B_0 e^{i(\phi-\omega t)}\,\hat{a}_z,= \omega \frac{E_0}{c} e^{i(\phi-\omega t)}\,\hat{a}_z
\end{equation}
therefore 
\begin{equation}
\vec{\nabla} \times \vec{E} = -\,\pdv{\vec{B}}{t}\quad\text{if and only if }\quad \omega={\frac{2c}{R_0}}.
\end{equation}
We have calculated both sides of Faraday's law from our ansatz and find that they are equal if and only if $\omega={2c}/{R_0}$. This sets the free parameter $\omega$. This renders the model monochromatic for a given $R_0$. There is \emph{only one} numerical value for $\omega$ which satisfies this law for a given $R_0$. The wave is \emph{monochromatic}. This is to be contrasted to plane waves which can have any given frequency. 

\subsubsection{Phase velocity}
We define our ansatz using the phase $\theta=(\phi - \omega t)$, representing the phase vortex of a circulating wave. The fitting parameter $\omega$ is determined by enforcing Faraday's law, yielding $\omega = 2c/R_0$. This constraint makes $\omega$ single-valued, resulting in a monochromatic wave. By definition, no group velocity exists since there is no wave packet with varying frequencies—for a given $R_0$, only one frequency is permitted.

We calculate the phase velocity to determine how the wavefront propagates along the ring. Expressing the phase in terms of arc-length $s = R_0\phi$ and time, we have $\theta(s,t) = s/R_0 - (2c/R_0)t$. The phase velocity corresponds to the speed at which points of constant phase move along the ring. Setting $d\theta = 0$ yields:

\begin{equation}
d\theta=\frac{1}{R_0}\,ds-\frac{2c}{R_0}dt=0
\quad\Rightarrow\quad
v_p = \frac{ds}{dt} = 2c
\end{equation}

\begin{equation}
v_p = 2c
\end{equation}

Since this represents phase velocity rather than group velocity, it does not violate special relativity. Superluminal phase velocities are common in waveguide structures, including toroidal geometries.

\subsection{Amp\`ere--Maxwell law}
We will use in the following form of Amp\`ere--Maxwell law 
\begin{equation}
 \vec{\nabla} \times \vec{B} -\frac{1}{c^2}\, \pdv{\vec{E}}{t}  =  \mu_0 \vec{J}.
\end{equation}

We use this unusual form of the law to emphasize that we will calculate the left side and right side independently. The left side, as can be trivially seen, can be derived from $\vec{E}$ and $\vec{B}$ by differentiation, while the right side can be derived independently from the continuity equation. 

The details are in appendix. The first derivation of $\vec{\nabla} \cdot \vec{J} $ from the $\vec{E}$ and $\vec{B}$ ansatz gives
\begin{equation}
\vec{\nabla} \cdot \vec{J} = \,\varepsilon_0 \omega \frac{E_0}{R_0} e^{i(\phi-\omega t)}.
\end{equation}
While calculating from the continuity equation and $\rho$ gives
\begin{equation}
\vec{\nabla} \cdot \vec{J} = -\,\pdv{\rho}{t}.
\end{equation}

yielding, given the $\rho$
\begin{equation}
\vec{\nabla} \cdot \vec{J} = \,\varepsilon_0 \omega \frac{E_0}{R_0} e^{i(\phi-\omega t)}.
\end{equation}

Which matches the $\vec{\nabla} \cdot \vec{J}$ quantity derived from the Amp\`ere--Maxwell law.

\subsection{A new class of solution for Maxwell equations}
This completes the proof that our ansatz satisfies Maxwell's equations, representing a novel solution to our knowledge. Unlike standard Hopfions or known vacuum cavity/waveguide eigenmodes, which are divergence-free, our solution possesses non-zero divergence in free space. We identify this divergence with the definition of charge density.

\section{Phenomenology of the electron}
Having established that our model accommodates charge, we investigate whether it reproduces QED electron phenomenology. We calculate RMS charge, spin, magnetic moment, and anomalous magnetic moment using parameter fitting.

\subsection{Calculation of the electron charge}
We have derived the charge density $\rho$ and we can integrate it over the volume of the torus to compute the total charge contained therein. 
\subsubsection{Space integrated total instantaneous charge inside the torus.}
The instantaneous charge density is 
\[
\rho(t) = \rho_0 \sin(\phi - \omega t)
\]
The real charge density, containing the phase factor, integrates to zero over the complete torus volume—positive and negative values cancel identically over the $2\pi$ spatial wavelength. This instantaneous charge clearly cannot represent the electron's phenomenological charge.  Here's a more concise version:

\subsubsection{Root Mean Square charge}

Since charge density forms a scalar wave, we adopt the RMS definition of charge. While total charge is well-defined for constant densities like the electron's $-e$, oscillating charge densities $\rho(\mathbf{r},t)$ require an effective measure. We employ the Root Mean Square (RMS) charge to quantify the effective magnitude of the fluctuating charge.

Computing the RMS charge over the volume gives us

\begin{equation}
 Q_{rms}= \sqrt{2}\pi^2\varepsilon_{0}\,E_{0}r_{0}^{2} =e
 \end{equation}

This provides a first constraint for our free parameters. 

\subsection{Magnetic Moment and gyromagnetic factor g}
Taking $\vec{J}$ given above, one can calculate the magnetic moment at any time as
\begin{equation}
  \vec{\mu} = \frac{1}{2}\int \vec{R} \times \vec{J}\,dV\,.
\end{equation}
A rather lengthy but straightforward derivation gives us a second constraint on our free parameters. We require that the magnetic moment be that of the electron and we write

\begin{equation}
\sqrt{2}\varepsilon_0 \pi\,c \,E_0 \,R_0 r_0^2  
\Big( 1 + \frac{r_0^2}{2R_0^2}\Big) =\mu_b (1+\frac{\alpha}{2\pi} +o(\alpha^2))
\end{equation}

With this form we emphasize the analytical expansion of the anomalous gyromagnetic calculated in QED in terms of powers of $\alpha$ with the first term being the Schwinger factor.

\subsection{Poynting Vector}

In order to calculate the Poynting vector we take the vector product between $\vec{E}$ and $\vec{B}$:
\begin{equation}
  \vec{S} \equiv \frac{1}{\mu_{0}}\vec{E}\times\vec{B},\qquad
\end{equation}

which gives us (details in appendix)
\begin{equation}
  \vec{S} = -\frac{1}{2}\,\varepsilon_{0}c\,E_{0}^{\,2}\,\hat{a}_{\phi},
\end{equation}

\subsection{Linear Momentum density}
Taking the Poynting vector, it is immediate to find the linear momentum 

\begin{equation}
  \vec{p} =\frac{\vec{S}}{c^2} = -\frac{1}{2}\,\frac{\varepsilon_{0}\,E_{0}^{\,2}}{c}\,\hat{a}_{\phi},
\end{equation}

\subsection{Angular Momentum and Electron Spin}

Taking $\vec{p}$ and the radius $R$ the angular momentum density is 
\begin{equation}
  \vec{l} = \vec{R}\times\vec{p}\,,
\end{equation}

The total angular momentum is 
\begin{equation}
  \vec{L} = \int \vec{R}\times\vec{p}\, dV
\end{equation}

Yielding (details in appendix) 
\begin{equation}
\vec{L}
= \frac{1}{c}\varepsilon_{0}\, E_{0}^{2}\, \pi^{2} R_0^2\, r_{0}^{2}
\left(1+ \frac{r_{0}^{2}}{4R_0^2}\right)\vec{a_z} 
\end{equation}

The spin is 1/2, means
\begin{equation}
\vec{L}= \frac{\hbar}{2}\,\hat{a}_{z}
\end{equation}

which is true if and only if 
\begin{equation}
\frac{1}{c}\varepsilon_{0}\, E_{0}^{2}\, \pi^{2} R_0^2\, r_{0}^{2}
\left(1+ \frac{r_{0}^{2}}{4R_0^2}\right) = \frac{\hbar}{2}\
\end{equation} 

Which provides our third constraint.

\subsection{Energy Density and mass}
We can now calculate the energy density.

\begin{equation}
u(t) = \frac{1}{2} \varepsilon_0 E^2(t) + \frac{1}{2\mu_0} B^2(t)
\end{equation}

Yielding energy density 
\begin{equation}
u(t) = 
\varepsilon_0E_0^2(1+\frac{R}{4R_0})
\end{equation}

Integrating the energy density over the volume of the torus we get the total energy contained therein (all details in appendix)

\begin{equation}
  U = \varepsilon_0 \pi^2 R_0 r_0^2E_0^2 (
  \frac{5}{2} +
  \frac{1}{8}\frac{r_0^2}{R_0^2})
\end{equation}

\subsection{Solving for $E_0, R_0, r_0$ and calculating $U$}

Using thin torus approximation $r_0/R_0 \approx 0$ and in first order in alpha $1+\alpha/2\pi \approx 1$; the equations simplify to

For spin: 
\begin{equation}
\frac{1}{c} \varepsilon_{0}E_{0}^{\,2}\pi^2 R_0^2r_0^2 
  = \frac{\hbar}{2}\
\end{equation} 

for charge: 
\begin{equation}
\sqrt{2}\pi^2\varepsilon_{0}\,E_{0}r_{0}^{2} =e
\label{eq:charge} 
\end{equation}

for magnetic moment:

\begin{equation}
\label{eq:mub}
\sqrt{2}\varepsilon_0 \pi \,c \,E_0 \,R_0 r_0^2  
 =\mu_b 
\end{equation}

For energy 
\begin{equation}
  U = \frac{5}{2}\varepsilon_0 \pi^2 R_0 r_0^2E_0^2
\end{equation}

Solving for $E_0$,  $R_0$, $r_0$ and $U$. We compare the amplitude to $E_S$, the Schwinger limit.  We compare lengths $R_0$ and $r_0$ to $r_c$ the reduced Compton wavelength.  Then we calculate total energy U and compare to mc2

\begin{equation}
E_0 = \frac{2\sqrt{2}}{\pi^2 } E_S \approx 0.2859 \, E_S \approx 3.783 \times 10^{17} \text{V/m}
\end{equation}

\begin{equation}
R_0 \approx \frac{\pi}{2} r_c  \approx 1.5726 r_c \approx 6.073 \times 10^{-13}  \text{m}
\end{equation}

\begin{equation}    
r_0 \approx \sqrt{\pi \alpha}  r_c \,\approx 0.1516 r_c \approx 5.854 \times 10^{-14} \text{m}
\end{equation}
\begin{equation}
U \approx \frac{5}{2 \pi} m_e c^2 \approx 0.7949 \,m_e c^2 \approx 0.406 \text{MeV}
\end{equation}

\subsection{Frequency $\omega$}
Using $\omega=\frac{2c}{R_0}$ and
$\frac{R_{0}}{r_{c}} \approx 1.573, $ we obtain 
\begin{equation}
\omega = 0.636\,\omega_D
\approx 9.86\times10^{20}\ \text{rad/s}.
\end{equation}
Where $\omega_D$ is the Dirac frequency (Zitterbewegung) from the Dirac equation.  

\subsection{Numerical values}
\begin{equation}
\boxed{
\frac{E_{0}}{E_{S}} \approx 0.286, \,
\frac{R_{0}}{r_{c}} \approx 1.573, \,
\frac{r_{0}}{r_{c}} \approx 0.152, \,
\frac{U}{m_{e}c^{2}} \approx 0.795, \,
\frac{\omega}{\omega_D} \approx 0.636}
\end{equation}

\subsection{Scwhinger limit for amplitude, Compton radius for dimensions}
This model predicts field amplitudes at the Schwinger limit, where electron-positron pair production occurs from vacuum. This correspondence supports our interpretation: the electron emerges as coiled light at precisely this critical amplitude. Furthermore, the characteristic scales match Compton dimensions, consistent with both experimental observations and the Dirac equation.

\subsection{On the Einstein mass energy equivalence}
While our result falls somewhat short of $m_ec^2$, it remains compelling: mass-energy equivalence emerges as the equivalence between mass and \emph{electromagnetic} energy. This energy is confined to a torus of Compton-scale dimensions with nearly the correct magnitude. The appearance of $\gamma$ in relativistic mass $E = \gamma m_e c^2,$ in classic special relativity further suggests an electromagnetic origin of mass, which emerges naturally from this framework.

\section{Conclusion}

We claim we have a novel microscopic representation of the QED electron. 

We hypothesized a circulating phase vortex ansatz with toroidal geometry, parameterized by major radius $R_0$, minor radius $r_0$, amplitude $E_0$, and frequency $\omega$. The phase contains unit prefactor for $\phi$.

Fitting to Maxwell's equations yielded geometrical charge $\rho$ from the field divergence, reversing the conventional ontological hierarchy between charges and fields. This constraint fixed $\omega$, rendering the wave monochromatic for given $R_0$.

Matching QED phenomenology required: RMS charge $Q_{RMS}=e$, magnetic moment $\mu=\mu_B(1+\alpha/2\pi)$, and spin $\hbar/2$.

The model yields field amplitudes at the Schwinger limit (where pair production occurs), Compton-scale dimensions for major radius, phase velocity $2c$, and Dirac time scale for the frequency, energy approximately $0.8m_ec^2$, all consistent with Dirac theory and observation. 
  
\section{Discussion}

\subsection{On the use of fitting parameters}

In using fitting parameters in any phenomenological approach one is well advised to remember the famous quip by John Von-Neumann, that “With four parameters I can fit an elephant, and with five I can make him wiggle his trunk.” L. Mayer later humorously formalized this joke in \cite{Mayer1993}, showing explicitly how to draw an elephant with four complex parameters (and wiggle its trunk with a fifth). This demonstration elevates the quip to the status of an aphorism, reminding us that, in general, overfitting models with too many free parameters, can reproduce almost anything, but the models lose their predictive power.

\subsection{Limits of the model}
We acknowledge this formulation as preliminary; notably, boundary conditions and far-field behavior cannot be satisfied with the Heaviside function formulation. The Heaviside mask turns off the fields at the boundary of the torus which is unphysical. Alternative representations employing Bessel functions may improve both accuracy and self-consistency. Current efforts are proceeding in this direction. This work is intended to provide a foundation for systematic refinement by the broader research community rather than constitute a definitive treatment.  The presented model nevertheless demonstrates encouraging agreement with QED phenomenology, yielding correct order-of-magnitude estimates for all parameters despite the oversimplified ansatz employed. While precise energy predictions were not anticipated given the elementary Heaviside step function implementation and the absence of proper boundary conditions for the torus, the results suggest that the approach captures essential physical features.

\subsection{A visualization and a realistic microscopic model}

We propose that this study validates pursuing Plain Old Electro-Magnetism (POEM) as a viable ontological framework for a microscopic understanding of the QED electron. Rather than replacing QED, the POEM electron provides crucial visualization capabilities, offering a classical microscopic electromagnetic framework that renders accessible to computation many esoteric quantum properties. Phenomena such as spin-1/2, the gyromagnetic moment and anomalous Scwhinger factor, and mass-energy equivalence—traditionally considered inherently quantum and beyond visualization—acquire straightforward visual explanations within this framework.

\subsection{A novel wave class}
This model's novelty lies in employing a fundamentally different class of wave solutions rather than plane waves. QED and condensed matter physics conventionally formulate creation and annihilation operators in terms of plane waves and photons, with Feynman diagram corrections requiring infinite summations of virtual photon contributions. We speculate that this infinite series may be a representation artifact arising from the mismatch between plane wave basis functions and the hypothesized underlying coiled wave structure—analogous to requiring infinite Fourier components to represent a square wave on a sine wave basis. 

\subsubsection{On Open Source}
The authors share this model in its preliminary form, embracing an open-source philosophy. Following the principles to "release early and release often" and "given enough eyeballs, all bugs are shallow," we present this work to attract collaborators with expertise in exotic electromagnetic envelope solutions. This simple model, developed with substantial AI assistance, seeks to engage researchers who can advance these ideas with greater proficiency in specialized electromagnetic formulations.
\subsection{Acknowledgments}
The authors thank Olivier Rousselle, Oliver Consa, and other colleagues from the Zitter Institute, for many insightful discussions. Peter Hagelstein, Emmanuel Fort, Herman Batelaan and John Bush for valuable feedback on early versions of the manuscript.

CAMS acknowledges the support of the INCT project Advanced Quantum Materials, involving the Brazilian agencies CNPq (Proc.408766/2024-7), FAPESP, and CAPES.
\

\bibliographystyle{Science}

\end{document}